\documentclass[conference]{IEEEtran}
\usepackage{ifpdf}

\usepackage[dvips]{graphicx}
\graphicspath{{../eps/}}
\DeclareGraphicsExtensions{.eps}

\usepackage{mathbbol}
\usepackage{cite}
\usepackage[T1]{fontenc} 
\usepackage{algorithm}
\usepackage{algorithmic}
\usepackage[cmex10]{amsmath}
\usepackage{amsthm}
\usepackage{amssymb}
\usepackage{balance}
\usepackage{eqparbox}
\usepackage{textcomp}
\usepackage{amsfonts}
\usepackage[eulergreek]{sansmath}
\usepackage{color}
\usepackage[cmintegrals]{newtxmath}

\newtheorem{theorem}{Theorem}
\newtheorem{remark}{Remark}
\usepackage[open,openlevel=1]{bookmark}
\usepackage{enumitem}
\usepackage{scalerel}
\usepackage{ctable}
\usepackage{threeparttablex}
\usepackage{scalerel}
\usepackage{mathtools}

\allowdisplaybreaks

\makeatletter
\newcommand\fs@betterruled{%
  \def\@fs@cfont{\bfseries}\let\@fs@capt\floatc@ruled
  \def\@fs@pre{\vspace*{8pt}\hrule height.8pt depth0pt \kern2pt}%
  \def\@fs@post{\kern2pt\hrule\relax}%
  \def\@fs@mid{\kern2pt\hrule\kern2pt}%
  \let\@fs@iftopcapt\iftrue}
\floatstyle{betterruled}
\restylefloat{algorithm}
\makeatother

\newtheorem{proposition}{Proposition}
\newcommand{\bs}[1]{\boldsymbol{#1}}

\newcommand{\mb}[1]{\mathbf{#1}}

\DeclareMathOperator*{\argmin}{arg\;min}
\DeclareMathOperator*{\argmax}{arg\;max}

\newcommand{\bseq}{\begin{subequations}}
	\newcommand{\eseq}{\end{subequations}}
\newcommand{\baln}{\begin{align}}
	\newcommand{\ealn}{\end{align}}
\newcommand{\balnd}{\begin{aligned}}
	\newcommand{\ealnd}{\end{aligned}}
\newcommand{\beq}{\begin{equation}}
	\newcommand{\eeq}{\end{equation}}
\newcommand{\beqn}{\begin{eqnarray}}
	\newcommand{\eeqn}{\end{eqnarray}}
\newcommand{\beqno}{\begin{eqnarray*}}
	\newcommand{\eeqno}{\end{eqnarray*}}
\newcommand{\bma}{\begin{displaymath}}
	\newcommand{\ema}{\end{displaymath}}
\newcommand{\bnu}{\begin{enumerate}}
	\newcommand{\enu}{\end{enumerate}}
\newcommand{\bce}{\begin{center}}
	\newcommand{\ece}{\end{center}}
\newcommand{\btb}{\begin{tabular}}
	\newcommand{\etb}{\end{tabular}}
\newcommand{\ba}{\begin{array}}
	\newcommand{\ea}{\end{array}}
\makeatletter
\newcommand{\superimpose}[2]{{%
  \ooalign{%
    \hfil$\m@th#1\@firstoftwo#2$\hfil\cr
    \hfil$\m@th#1\@secondoftwo#2$\hfil\cr
  }%
}}
\makeatother

\begin{document}
\pagenumbering{gobble}

%

\title{User-Centric Beam Selection and Precoding Design for Coordinated Multiple-Satellite Systems \vspace{-2mm}}

 \author{\IEEEauthorblockN{Vu Nguyen Ha${}^{\dagger}$, 	
Duy H. N. Nguyen${}^{\ddagger}$, Juan C.-M. Duncan${}^{\dagger}$, Jorge L. Gonzalez-Rios${}^{\dagger}$, Juan A. Vasquez${}^{\dagger}$,  \\  Geoffrey Eappen${}^{\dagger}$, Luis M. Garces-Socarras${}^{\dagger}$,  Rakesh Palisetty${}^{\flat}$, Symeon Chatzinotas${}^{\dagger}$, and Bj\"{o}rn Ottersten${}^{\dagger}$
}
		\IEEEauthorblockA{${}^{\dagger}$\textit{Interdisciplinary Centre for Security, Reliability and Trust (SnT), University of Luxembourg, Luxembourg}\\
  \textit{${}^{\ddagger}$Department of Electrical and Computer Engineering, San Diego State University, San Diego, CA, USA 92182}\\
  \textit{${}^{\flat}$Shiv Nadar Institution of Eminence, Delhi NCR, Greater Noida 201314, India.}}
  \vspace{-8mm}
   }

	
	

	\maketitle
	
		\begin{abstract}
			This paper introduces a joint optimization framework for user-centric beam selection and linear precoding (LP) design in a coordinated multiple-satellite (CoMSat) system, employing a Digital-Fourier-Transform-based (DFT) beamforming (BF) technique. Regarding serving users at their target SINRs and minimizing the total transmit power, the scheme aims to efficiently determine satellites for users to associate with and activate the best cluster of beams together with optimizing LP for every satellite-to-user transmission. These technical objectives are first framed as a complex mixed-integer programming (MIP) challenge. To tackle this, we reformulate it into a joint cluster association and LP design problem. Then, by theoretically analyzing the duality relationship between downlink and uplink transmissions, we develop an efficient iterative method to identify the optimal solution. Additionally, a simpler duality approach for rapid beam selection and LP design is presented for comparison purposes. Simulation results underscore the effectiveness of our proposed schemes across various settings.
		\end{abstract}
		
		\begin{IEEEkeywords}
			Multibeam SATCOM, Linear Precoding, Coordinated Multiple Satellites, Satellite Association, Beam Selection.
    \vspace{-5mm} 
	\end{IEEEkeywords}
	
	

	\section{Introduction}
\vspace{-1mm}

Satellite communications (SATCOM) with dense satellite deployment have been considered as an important radio interface that can fulfill global coverage and high data-rate demands of the next-generation wireless networks \cite{NTNSurvey22}. 
Launching a huge number of spacecrafts on different orbits to form mega constellations can help future SATCOM systems enlarge the coverage and signal strength by reducing the satellite-user distances. 
Additionally, LP \cite{Angeletti_Access20} and coordinated multi-point transmission (CoMP) \cite{VuHa_TVT16} technologies for a group of neighboring satellites can be exploited to further enhance the network capacity \cite{kim2023coverage}. However, 
employing these advanced technologies also poses challenges in mitigating the 
inter-beam and cross-coverage interference. To tackle this, one needs to evolve LP and CoMP designs in the context of dense satellite deployment \cite{vazquez2016precoding}.

Optimizing interference avoidance in multi-beam (MB), multi-satellite (MSat) systems is crucial. 
To realize the MB transmission, the European Space Agency (ESA) has proposed low-complexityBF algorithms using DFT codebooks for massive multi-input multi-output (mMIMO) payloads to significantly boost network throughput \cite{Angeletti_Access20}.
The study in~\cite{you2020massive} introduced an mMIMO approach for low-earth orbit (LEO) satellites (LEOSats), utilizing full-frequency-reuse downlink precoding and uplink detection based on statistical channel state information (CSI) to enhance
the signal-to-noise ratio (SINR). Further, hybrid precoding frameworks for mMIMO SATCOM have been proposed in~\cite{Liu_Access22}, yet their computational demands challenge payload implementation.
The work in \cite{kim2023coverage} aimed at exploiting the advantages of cooperative BF from a cluster of neighboring LEOSats to improve the network capacity and coverage.
In \cite{Hongtao_TVT23}, authors proposed to utilize flexible beam scheduling and BF of MB LEOSats to increase the accuracy of a positioning system by sending signals simultaneously from different aircrafts.
Authors in \cite{roper2022distributed} proposed a novel downlink BF approach for a swarm of LEOSats which can massively increase the network spectral efficiency.
However, implementing the cooperative transmission for MSat systems has to deal with a critical synchronization challenge.

 
 In this work, we focus on developing an innovative optimization framework to transform user-centric beam selection and LP within a CoMSat system, utilizing DFT BF. 
 Designed to minimize transmission power, our method accelerates satellite selection for user connections, identifies the best beam clusters, and optimizes LP for serving users at their target SINRs efficiently. Initially posed as a complex MIP-tacking task, we simplify this into a joint problem of cluster association and LP design.
A key aspect of our research is analyzing the duality between downlink and uplink transmissions, leading to an iterative process that accurately identifies the best solutions.
We also offer a simpler, duality-based method for beam selection and LP design as a point of comparison. Our simulations validate the efficiency and flexibility of our approaches, demonstrating marked performance enhancements across different scenarios. By pushing the boundaries of SATCOM technology, our work lays the groundwork for more efficient strategies in the future.

\vspace{-1mm}

  \section{System Model and Problem Formulation} 
  \vspace{-1mm}\label{sec:SMnPF}
	\subsection{System Model}
\begin{figure}[!t]
		\centering
		\includegraphics[width=70mm]{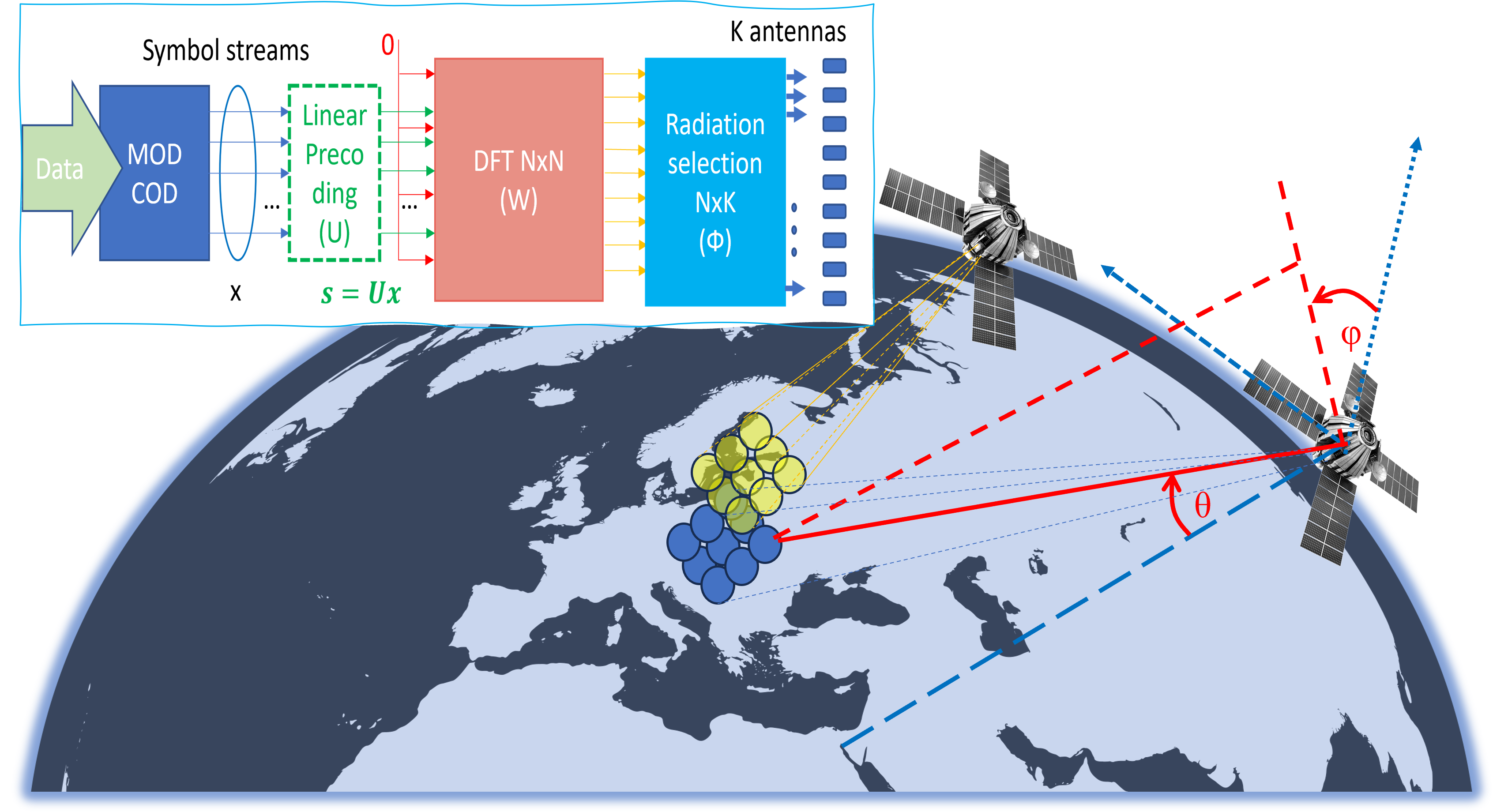}
		\vspace{-4mm}
		\caption{A coordinated MB MSat system with full potential beams.}
		\label{Pload_Archi}
		\vspace{-3mm}
	\end{figure}

\begin{figure}[!t]
		\centering
	\includegraphics[width=70mm]{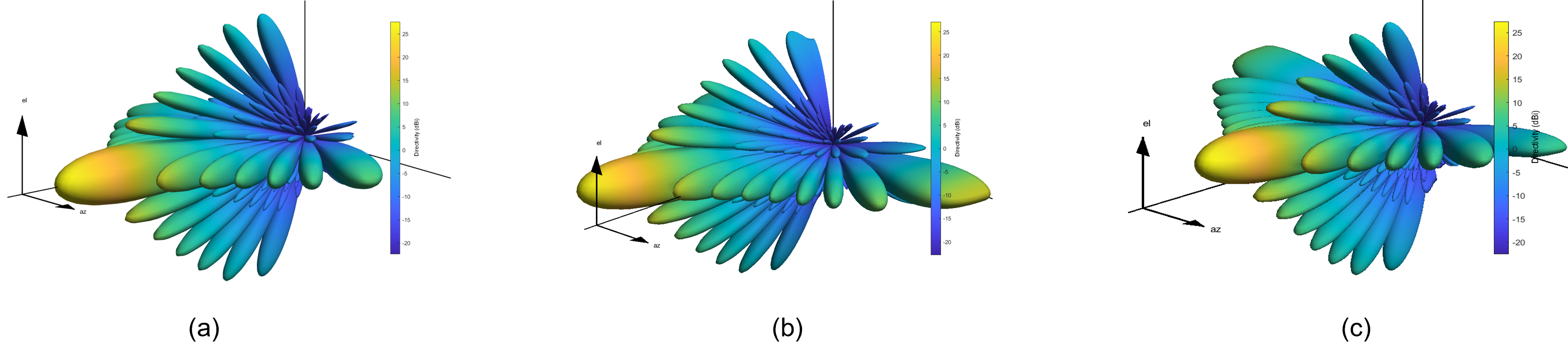}
 \vspace{-4mm}
		\caption{Examples of beam pattern due to selected DFT-vector for $N = 256$: (a) DFT-vector $\mb{w}_1$, (b) DFT-vector $\mb{w}_5$, and (c) DFT-vector $\mb{w}_{25}$ \cite{VuHa_GCWS22}.}
		\label{Beam_pattern_example}
		\vspace{-4mm}
	\end{figure}

\begin{figure}[!t]
    \centering
    \includegraphics[width=80mm]{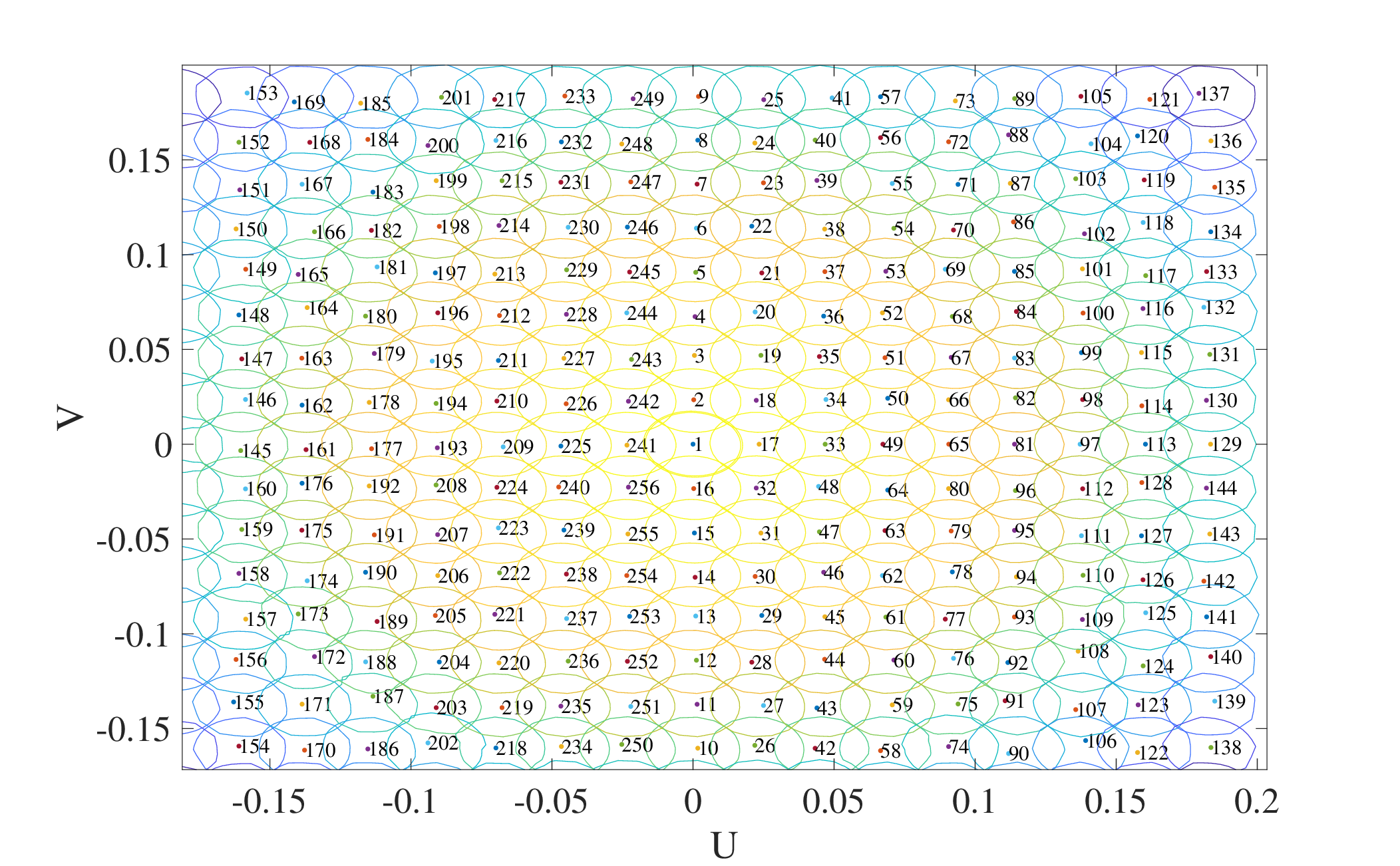}
    \vspace{-4mm}
    \caption{$3$-dB footprints of $256$ beams using $100$ sub-array antennas \cite{Rakesh_PIMRC23}.}  \label{fig:SNR8x8taper}
    \vspace{-5mm}
\end{figure}

	This paper considers a DFT-BF-enabled LEO constellation consisting of $L$ LEOSats serving $M$ ground single-antenna users, as illustrated in Fig.~\ref{Pload_Archi}. 
    One assumes that every satellite is equipped with a uniform rectangular array (URA) of $K$ antennas based on which the DFT-BF technology is employed to create MB transmission as discussed in \cite{VuHa_GCWS22}. 
    We considered the users to be very small-aperture terminals (VSATs) with perfect tracking mechanisms, therefore with a constant receiver's antenna gain.
    Denote $N$ the size of the DFT-BF vectors, examples of beam patterns corresponding to various DFT vectors and $3$-dB beam footprints with $N = 16 \times 16$ and $K = 10 \times 10$ are illustrated in Figs.~\ref{Beam_pattern_example} and \ref{fig:SNR8x8taper}, respectively, where the $[U,V]$ coordinates given in Fig.~\ref{fig:SNR8x8taper} are defined as $U^{\ell}_m=\sin(\theta_m^{\ell})\cos(\phi_m^{\ell})$ and $V^{\ell}_m=\sin(\theta_m^{\ell})\sin(\phi_m^{\ell})$ \cite{Angeletti_Access20}. 
 
In this scheme, various LEOSats can be selected to communicate to different users.
To address this LEO-user association, we introduce $ a^{\ell}_m$ as the matching binary variables, i.e.
$ a^{\ell}_m = 1$ if satellite $\ell$ is assigned to serve user $m$ and 
$ a^{\ell}_m = 0$ otherwise.
Due to the synchronization challenge, the cooperative transmission from multiple LEOSats to each user has not been considered in this work. In this context, only one LEOSat is selected to serve any user at a specific time, which yields 
\beq
\vspace{-1mm} 
(C1): \quad \scaleobj{.8}{\sum_{\forall \ell}} a^{\ell}_m = 1 \quad \forall m.
\vspace{-1mm} 
\eeq
 
\subsubsection{DFT-Effective CSI}
Assume the same DFT-BF design is employed at all the payloads. To ease the notation, we denote beam $n$ the propagation pattern due to DFT-vector $n$, denoted as $\mb{w}^{\sf{DFT}}_n \in \mathbb{C}^{K \times 1}$.
Let $\mb{h}^{\ell}_m \in \mathbb{C}^{K \times 1}$ be the channel vector from LEOSat $\ell$ to user $m$.
If user $m$'s data is transmitted over beam $n$ of LEOSat $\ell$, its signal will be multiplied with $\mb{w}^{\sf{DFT}}_n$ before being propagated through all antennas \cite{VuHa_GCWS22,Rakesh_PIMRC23}.
Hence, the DFT-effective channel coefficient to user $m$ from beam $n$ of LEOSat $\ell$ can be denoted as
\beq
g^{\ell}_{n,m} = \mb{h}^{\ell,H}_m \mb{w}_n^{\sf{DFT}}.
\eeq
\subsubsection{Beam-wise Linear Precoding}
As illustrated in Figs.~\ref{Beam_pattern_example} and \ref{fig:SNR8x8taper}, applying different DFT vectors can result in beam patterns with different pointing directions. Hence, $g^{\ell}_{n,m}$ can be sufficiently strong or neglectable according to the location of user $m$ from LEOSat $\ell$'s point of view.
In addition, to enhance MB multiplexing transmission, one can employ beam-wise LP vectors to different clusters of beams to serve all users. 

Regarding this beam-wise LP design, one denotes $u^{\ell}_{n,m} \in \mathbb{C}$ as the LP coefficient corresponding signal of user $m$ and beam $n$ of satellite $\ell$.
This LP coefficient can be applied to the baseband signal transmitted to user $m$, which will be propagated by all antennas with weights defined by DFT vector $\mb{w}_n$.
Here, the beam selection to serve user $m$ can be cast by norm-$\ell_0$, i.e., $\| u^{\ell}_{n,m}\|_0 = 1$ if beam $n$ is assigned to serve user $m$, and $\| u^{\ell}_{n,m}\|_0 = 0$ otherwise.
Due to the limited computation power, one assumes that every cluster formed to serve any user at each LEOSat contains at most $B$ beams. This constraint can be written as
\beq
(C2): \quad \scaleobj{.8}{\sum_{\forall n}} \| u^{\ell}_{n,m}\|_0 \leq B \text{ if } a^{\ell}_m = 1, \quad \forall (\ell,m). 
\eeq

The cluster serving each user can be formed by carefully selecting beams from a predetermined set of beams having efficient links to that user. These beams can be predetermined based on the location of users and the beams' footprint given in Fig.~\ref{fig:SNR8x8taper}.
Denote $\mathcal{B}^{\ell}_m$ as such set corresponding to user $m$ at satellite $\ell$.
Regarding the beam-wise LP and DFT-effective CSI, the received signal at user $m$ can be expressed as
	\beqn \label{eq:resig}
	y_m & = & {\scaleobj{.8}{\sum_{\forall \ell}}} \mb{h}^{\ell,H}_m \scaleobj{.8}{\sum_{j \in \mathcal{M}} \sum_{n \in \mathcal{B}^{\ell}_j}} a^{\ell}_j \mb{w}^{\sf{DFT}}_n u^{\ell}_{n,j} x_j + \eta_m \nonumber \\
 & = & {\scaleobj{.8}{\sum_{\forall \ell} \sum_{j \in \mathcal{M}} \sum_{n \in \mathcal{B}^{\ell}_j}}} a^{\ell}_j g_{n,m}^{\ell} u^{\ell}_{n,j} x_j + \eta_m,
	\eeqn
where $\mathcal{M}$ denotes the set of all users and $\eta_m$ is the additive noise. Thus, the SINR of user $m$ can be given as
	\beq
	\Gamma_m(\mb{U},\mb{A}) = \dfrac{\sum_{\forall \ell} a^{\ell}_m \sum_{n \in \mathcal{B}^{\ell}_m} \vert g_{n,m}^{\ell} u^{\ell}_{n,m}\vert^2}{\sum_{j \neq m} \sum_{\forall \ell} a^{\ell}_j \sum_{n \in \mathcal{B}^{\ell}_j} \vert g_{n,m}^{\ell} u^{\ell}_{n,j} \vert^2 + \sigma^2}\,,
	\eeq
 where $(\mb{U},\mb{A})$ are matrices representing all $u_{u,m}$'s and $a^{\ell}_m$'s while $\sigma^2$ denotes the noise power.


\subsection{Problem Formulation}
Our design aims to jointly optimize satellite-user association, determining the beam cluster of the assigned LEOSat serving each user, and determining the corresponding beam-wise LP coefficients to minimize the total power transmission under the constraints on users' QoS. 
Then, these technical design objectives can be cast by the following power minimization problem,
	\begin{subequations} \label{obj_1}
		\begin{eqnarray} 
(\mathcal{P}_1) & \min \limits_{\mb{A},\mb{U}} & {}\scaleobj{.8}{\sum \limits_{\forall \ell} \sum \limits_{\forall m} \sum \limits_{n \in \mathcal{B}^{\ell}_m}} \vert u^{\ell}_{n,m} \vert^2   \\
& \text{s.t.} & \text{ constraints $(C1), (C2)$}, \nonumber \\
& & (C3): \;\; \Gamma_m(\mb{U},\mb{A}) \geq \bar{\gamma}_m, \;\;\forall m \in \mathcal{M},
\end{eqnarray} 
\end{subequations}
where $(C3)$ represents the the required QoS of 
users.
As can be seen, $(\mathcal{P}_1)$ is a mixed integer programming which is well-known as NP-hard and very challenging to solve.
To deal with this critical issue, we propose 
a novel low-complexity algorithm in the following.

\section{Problem Formulation and Strong Duality}
\vspace{-1mm} 
\subsection{Cluster-Association-based Reformulation}
\vspace{-1mm}
To address the challenges in solving $(\mathcal{P}_1)$ due to the binary variables and the non-linear function of the sparsity term of LP coefficients, we first consider the following remark.
\vspace{-1mm} 
\begin{remark} \label{rmk2}
Regarding constraint $(C2)$, one can observe that:
    \begin{enumerate}
        \item[(a)] If an optimal cluster solution for any user $m$ from LEOSat $\ell$ contains less than $B$ beams, we add some others from $\mathcal{B}^{\ell}_m$ to obtain a $B$-beam cluster and set the LP coefficients corresponding to these added beams to zeros.
        \item[(b)] There are a finite number of potential clusters containing $B$ beams which can be generated from any $\mathcal{B}^{\ell}_m$. Instead of dealing with sparsity terms in $(C2)$, one can hence tend to select the efficient one within these potential clusters.
    \end{enumerate}
\end{remark}
\vspace{-1mm} 

Thanks to Remark~\ref{rmk2}, $(\mathcal{P}_1)$ can be re-formed to a joint cluster association (CA) and LP design problem as follows.
Let $\mathcal{S}^{\ell}_m$ be the set of all possible $B$-beam clusters created from $\mathcal{B}^{\ell}_m$. 
Here, each member of $\mathcal{S}^{\ell}_m$ is formed from $B$ beams of $\mathcal{B}^{\ell}_m$. 
It is worth noting that when $\mathcal{B}^{\ell}_m \neq \varnothing$ and $|\mathcal{B}^{\ell}_m| \leq B$, $\mathcal{S}^{\ell}_m$ can be set to contain only one cluster which is formed by all beams in $\mathcal{B}^{\ell}_m$.
Then, if $|\mathcal{B}^{\ell}_m| > B$, number of possible clusters from LEOSat $\ell$  can be defined as
\vspace{-2mm}

\beq
T^{\ell}_m = \scaleobj{.8}{{B^{\ell}_m \choose B}}=\scaleobj{.8}{\dfrac{B^{\ell}_m!}{B!(B^{\ell}_m-B)!}},
\vspace{-1mm}
\eeq
where $B^{\ell}_m = |\mathcal{B}^{\ell}_m|$. 
Here, one can recall that $T^{\ell}_m = 0$ if $\mathcal{B}^{\ell}_m = \varnothing$ and $T^{\ell}_m = 1$ if $\mathcal{B}^{\ell}_m \neq \varnothing$ and $|\mathcal{B}^{\ell}_m| \leq B$.
We further denote $\mathcal{S}_m$ the set of all possible $B$-beam clusters from all LEOSats corresponding to user $m$, i.e.,
$\mathcal{S}_m = \bigcup_{\forall \ell} \mathcal{S}^{\ell}_m$.
\vspace{-1mm} 

\begin{remark}
    From $\mathcal{S}_m$, the tasks of beam selection and satellite assignment in $(\mathcal{P}_1)$ can be cast by CA for each user.
Specifically, once a cluster from $\mathcal{S}_m$ is selected to serve user $m$, the LEOSat and beams corresponding to this cluster can be considered as the solution for user $m$ in problem $(\mathcal{P}_1)$.
\end{remark}
\vspace{-1mm} 

Denote $T_m = \vert \mathcal{S}_m \vert = \sum_{\forall \ell} T^{\ell}_m$.
Then, any specific cluster in $\mathcal{S}_m$, denoted as $\mathcal{C}_m^t$ ($1 \leq t \leq T_m$), can be chosen to serve user $m$. 
Let $\mb{u}_{m}^t = \left[u_{n,m} | n \in \mathcal{C}^t_m \right]$ denote a vector formed by beam-wise LP coefficients serving user $m$ of all beams in cluster $\mathcal{C}_m^t$. 
We also denote $\mb{g}_{j,m}^t$ the concatenated channel vector from beams in $\mathcal{C}_m^t$ to user $j$ which is generated from all $g^{\ell}_{n,j}$ of LEOSat $\ell$ and all beam $n$ corresponding to $\mathcal{C}^t_m$.
Then, the SINR of user $m$ if it is served by cluster $\mathcal{C}_m^t$ can be given as
\vspace{-3mm} 

\beq
\hat{\Gamma}_m^t = { \vert  \mb{g}_{m,m}^{t H} \mb{u}_m^t\vert^2 }/{\scaleobj{.8}{\sum \limits_{j \in \mathcal{M}/m} \sum \limits_{k \in \mathcal{S}_j}}  \vert \mb{g}_{m,j}^{k H} \mb{u}_j^k \vert^2 + \sigma^2 }.
\eeq
\vspace{-3mm} 

\noindent
Let binary variable $\lbrace r_m^t \rbrace$, ($m \in \mathcal{M}$, $1 \leq t \leq T_m$) represent the association between user $m$ and cluster $\mathcal{C}_m^t$, i.e., $ r_m^t = 1$ if  user $m$ is served by cluster $\mathcal{C}_m^t$, $ r_m^t = 0$ otherwise.
Then, the joint cluster-associate and LP design problem is defined as 
	\begin{subequations} \label{obj_2}
		\begin{eqnarray} 
 \hspace{-10mm}  (\mathcal{P}_2) &\min \limits_{\lbrace \mathbf{u}_m^t \rbrace, \lbrace r_m^t \rbrace} & \hspace{-2mm} \Phi(\mb{U},\lbrace r_m^t \rbrace) = \scaleobj{.8}{\sum \limits_{m \in \mathcal{M}} \sum \limits_{t \in \mathcal{S}_m}} \mb{u}_m^{tH} \mb{u}_m^t    \\
& \text{s.t.} & \hspace{-5mm} (C3'):\; \scaleobj{.8}{ \sum \limits_{t \in \mathcal{S}_m}}r_m^t \hat{\Gamma}_m^t  \geq \bar{\gamma}_m, \forall m \in \mathcal{M}, \\
& {} & \hspace{-5mm} (C4): \; \scaleobj{.8}{\sum_{\forall t}}r_m^t =1, \forall m \in \mathcal{M}, 
\end{eqnarray} 
\end{subequations}
where 
 $(C4)$ guarantees that every user is associated with only one cluster. 
Unfortunately, $(\mathcal{P}_2)$ is still a challenging MIP.

\subsection{Joint Cluster-Association and LP Design}
\vspace{-1mm}
\subsubsection{Binary-Variable Omitting Transformation}
It needs to be noticed that $r_j^k = 0$ if $\mb{u}_j^{kH}\mb{u}_j^k = 0$ and vice versa. Based on this, a novel algorithm is developed to deal with problem $(\mathcal{P}_2)$ by first omitting $\lbrace r_m^t \rbrace$ to form the following problem,
\begin{subequations} \label{obj_3}
		\begin{eqnarray} 
 \hspace{-10mm}  (\mathcal{P}_3) &\min \limits_{\lbrace \mathbf{u}_m^t \rbrace, \lbrace r_m^t \rbrace} & \hspace{-2mm} \Phi(\mb{U},\lbrace r_m^t \rbrace) = \scaleobj{.8}{\sum \limits_{m \in \mathcal{M}} \sum \limits_{t \in \mathcal{S}_m}} \mb{u}_m^{tH} \mb{u}_m^t   \\
\hspace{-10mm} & \text{s.t.} & 
\hspace{-5mm} (\tilde{C3}'):\; \scaleobj{.8}{ \sum \limits_{t \in \mathcal{S}_m}}\hat{\Gamma}_m^t  \geq \bar{\gamma}_m, \forall m \in \mathcal{M}.
\end{eqnarray} 
\end{subequations}
\vspace{-5mm}

\begin{proposition} \label{lm2}
The relationship between $(\mathcal{P}_2)$ and $(\mathcal{P}_3)$ can be cast in the following comments:
\begin{enumerate}
    \item[i)] The minimum total transmit power
obtained from solving $(\mathcal{P}_3)$
is a lower-bound to that obtained from solving $(\mathcal{P}_2)$.
\item[ii)] Let $\lbrace \mb{u}_m^{t \ast} \rbrace$ be an optimal solution of (\ref{obj_3}). Then,  $\lbrace \mb{u}_m^{t \ast} \rbrace$ is also a feasible solution for $(\mathcal{P}_2)$ if 
``there is only one $t$, $1 \leq t \leq T_m$ for each user $m$ such that $\mb{u}_m^{t \ast H} \mb{u}_m^{t \ast} > 0$ and $\mb{u}_m^{k \ast H} \mb{u}_m^{k \ast} = 0$, $\forall k \neq t$''. This ``IF'' is named as \textbf{Cond-1}.
\end{enumerate}
\end{proposition}
\vspace{-1mm}

\begin{IEEEproof} The proposition can be proved briefly as follows:

\noindent
i) Regarding that that $r_m^t = 0$ if $\mb{u}_m^{tH}\mb{u}_m^t = 0$ and vice versa, the optimal LP solution of $(\mathcal{P}_2)$ must satisfy constraint $(\tilde{C3}')$ of $(\mathcal{P}_3)$. Hence, the minimum total transmit power obtained from solving problem $(\mathcal{P}_3)$ is a lower-bound to that obtained from solving problem $(\mathcal{P}_2)$.

\noindent
ii)  Then, if $\lbrace \mb{u}_m^{t \ast} \rbrace$ satisfies \textbf{Cond-1}, $\lbrace \mb{u}_m^{t \ast} \rbrace$ must be also a feasible solution for $(\mathcal{P}_2)$. Therefore, $\lbrace \mb{u}_m^{t \ast} \rbrace$ satisfying \textbf{Cond-1} must be an optimal solution the problem $(\mathcal{P}_2)$. 
\end{IEEEproof}
Thanks to Proposition~\ref{lm2}, it is left to find a solution of $(\mathcal{P}_3)$ that should satisfy \textbf{Cond-1}.
 
\subsubsection{Strong Duality $(\mathcal{P}_3)$}
We first express the Lagrangian function of problem $(\mathcal{P}_3)$ as
\vspace{-5mm}

\beqn \label{eq:Lagr}
& \hspace{-4mm}\mathcal{L}(\mb{U},\boldsymbol{\lambda}) & \!\!\!\!\! = \!\!\! \scaleobj{.8}{\sum \limits_{\forall (m,t)}} \!\! \mb{u}_m^{tH}\mb{u}_m^t \!\! - \!\!\!
 \scaleobj{.8}{\sum \limits_{m \in \mathcal{M}}} \!\! \lambda_m \!\Big( \scaleobj{.8}{\sum \limits_{t \in \mathcal{S}_m}  \dfrac{\left|  \mb{g}_{m,m}^{tH} \mb{u}_m^t\right| ^2}{\bar{\gamma}_m} \! - \!\! \sum \limits_{j \neq m} \sum \limits_{k \in \mathcal{S}_j}  \left| \mb{g}_{m,j}^{kH} \mb{u}_j^k \right|^2 \! - \sigma^2} \Big)  \nonumber \\
&& \!\!\!\!\! =\!\!\! \scaleobj{.8}{\sum \limits_{m \in \mathcal{M}}} \lambda_m \sigma^2 + \scaleobj{.8}{\sum \limits_{m \in \mathcal{M}} \sum \limits_{t \in \mathcal{S}_m}} \mb{u}_m^{tH} \Omega_m^t(\boldsymbol{\lambda}) \mb{u}_m^t,
\eeqn
\vspace{-3mm}

\noindent
where $\Omega_m^t(\boldsymbol{\lambda}) = \mb{I}_m^t - \scaleobj{0.8}{\dfrac{\lambda_m}{\bar{\gamma}_m}} \mb{g}_{m,m}^t\mb{g}_{m,m}^{tH}  + \scaleobj{1}{\sum_{j \neq m} } \lambda_j \mb{g}_{j,m}^t\mb{g}_{j,m}^{tH}$, $\lambda_m$'s are Lagrange multipliers of $(\tilde{C3}')$ and 
$\boldsymbol{\lambda}=[\lambda_1,...,\lambda_M]$. Regarding \eqref{eq:Lagr}, the dual 
function can be given as
\vspace{-4mm}

\beq
\mathsf{g}(\boldsymbol{\lambda}) = \min_{\mb{U}} \mathcal{L}(\mb{U},\boldsymbol{\lambda}).
\eeq 
\vspace{-5mm}

\noindent
It can be verified from (\ref{eq:Lagr}) that if any matrix $\Omega_m^t(\boldsymbol{\lambda})$ is not positive semi--definite, there will exist $\mb{u}_m^t$ that makes $g(\boldsymbol{\lambda},\mu)$ unbounded below\footnote{$\mathbf{A} \succ \mathbf{0}$, $\mathbf{A} \succeq \mathbf{0}$, and $\mathbf{A} \, {\overset{\nsucc}{\textunderscore}} \, \mathbf{0}$ indicate that $\mathbf{A}$ is positive definite, positive semi--definite, and positive semi--definite but not positive definite, respectively.}.
Then, the dual problem can be written as
\vspace{-4mm}

\begin{subequations} \label{obj_5}
\begin{eqnarray}
\hspace{-5mm} (\mathcal{P}^{\sf{D}}_3) & \max \limits_{\boldsymbol{\lambda} \geq 0} & {} \scaleobj{.8}{\sum \limits_{m \in \mathcal{M}}} \lambda_m \sigma^2 \\
& \hspace{-18mm} \text{s.t.} &  \hspace{-12mm} (C5)\!: \mb{I}_m^t  \! + \!\! \scaleobj{.8}{\sum \limits_{j \neq m} } \lambda_j \mb{g}_{j,m}^t\mb{g}_{j,m}^{tH}\succeq \scaleobj{.8}{\dfrac{\lambda_m}{\bar{\gamma}_m}} \mb{g}_{m,m}^t\mb{g}_{m,m}^{tH}, \forall (t,m). \label{cnt_C5}
\end{eqnarray} 
\end{subequations}
Note that when $(\mathcal{P}_3)$ is infeasible, one has a weak duality between the primary and dual problems. Then, the dual problem $(\mathcal{P}^{\sf{D}}_3)$ will be unbounded above, i.e., $\lambda_m \rightarrow \infty$. 
Hence, in this work, we assume that problem $(\mathcal{P}_3)$ is feasible.

\begin{theorem} \label{thm_duality}
The strong duality holds between $(\mathcal{P}_3)$ and $(\mathcal{P}^{\sf{D}}_3)$ if problem $(\mathcal{P}_3)$ is feasible.
\end{theorem}

\begin{IEEEproof}
Denote $\bs{\lambda}^{\star}$ as the optimal solution of $(\mathcal{P}_3)$. 
Thanks to the strong duality properties given in \cite{HoangTuy_JGO13,DuyNguyen_TWC17}, to fulfill the proof, it is left to find $\mb{U}^{\star} = \argmin_{\mb{U}} \mathcal{L}(\mb{U},\bs{\lambda}^{\star})  $ which is also a feasible solution of $(\mathcal{P}_3)$ and satisfies the following,
\beq \label{lagrangian_cnt}
\lambda_m \Big( \scaleobj{.8}{\sum \limits_{\forall t}  {\left|  \mb{g}_{m,m}^{tH} \mb{u}_m^t\right| ^2}/{\bar{\gamma}_m}  -  \sum \limits_{j \neq m} \sum \limits_{\forall k}  \left| \mb{g}_{m,j}^{kH} \mb{u}_j^k \right|^2  - \sigma^2} \Big) =  0, \; \forall m.
\eeq
We will prove this by employing the contradiction approach. Assume there exists $m \in \mathcal{M}$ such that $\Omega_{m}^{t}(\boldsymbol{\lambda}^{\star}) \succ \mathbf{0}$ for all $t$. Then, we can keep all $\lambda_j^{\star}$ ($j \neq m$) unchanged, and increase $\lambda_m^{\star}$ to a value $\bar{\lambda}_m$ corresponding to which there appears $t_m$ with holding the corresponding constraint~\eqref{cnt_C5}, (i.e. $\overset{\nsucc}{\textunderscore} \, \mathbf{0})$. 
Hence, we can find another feasible solution of $(\mathcal{P}^{\sf{D}}_3)$ and $\sum_{\forall m} \lambda_m > \sum_{\forall m} \lambda_m^{\star}$, which results in a contradiction.
Hence, for each user $m$, one can determine at least one $t_m$ so that $\Omega_{m}^{t_m}(\boldsymbol{\lambda}^{\star}) \, {\overset{\nsucc}{\textunderscore}} \, \mathbf{0}$.
For this cluster of user $m$, there exists $\hat{\mathbf{u}}_{m}^{t_m}$ such that $\hat{\mathbf{u}}_{m}^{t_m H} \Omega_{m}^{t_m} \hat{\mathbf{u}}_{m}^{t_m} =0.$
According to $\hat{\mathbf{u}}_{m}^{t_m}$, $\mathbf{u}_{m}^{t_m} \neq \mb{0}$ can be defined.
Then, all other clusters $t$, $t \in \mathcal{S}_m/\{t_m\}$, with strict inequalities, i.e. $\Omega_{m}^{t}(\boldsymbol{\lambda}^{\star}) \succ \mathbf{0}$, we can set $\mathbf{u}_{m}^{t \star}$ to all-0 vector. 
\end{IEEEproof}

\begin{remark} \label{lm3}
The proof has also suggested an efficient approach to obtain $\bs{\lambda}^{\star}$ by alternatively updating one element of $\bs{\lambda}$. At the convergence point, the probability of having two or more clusters $t_m$ at a specific value of $\lambda_m$ that have that $\Omega_{m}^{t_m}(\boldsymbol{\lambda}^{\star}) \, {\overset{\nsucc}{\textunderscore}} \, \mathbf{0}$, is almost zero for most practical systems, except for the cases where the channels from two groups are exactly symmetric.
Hence, it almost surely happens that all other constraints~\eqref{cnt_C5} are strict, i.e.,
$\Omega_{m}^{t}(\boldsymbol{\lambda}^{\star}) \succ \mathbf{0}, \forall t \neq t_m$. 
Regarding this, the optimal solution of $(\mathcal{P}_3)$ should satisfy \textbf{Cond-1}.
\end{remark} 

\section{Downlink-Uplink Duality-based LP Designs} 


\subsection{Dual Uplink System Model}
We examine a dual virtual uplink system where each cluster acts as a base station. The uplink channel matrices are obtained by transposing the downlink ones, assuming the noise at cluster $\mathcal{C}^t_m$ follows a zero-mean AWGN with covariance matrix $\sigma^2\mathbf{I}$. Let $\hat{p}_m$ represent the transmitting power of user $m$. When user $m$ is associated with cluster $\mathcal{C}^t_m$, the system employs the receiving BF vector $\hat{\mb{u}}^t_m$ to decode the signal of user $m$.
In this uplink scenario, the CA involves selecting a cluster from $\mathcal{S}_m$ that allows user $m$ to meet the SINR target $\bar{\gamma}_m$ with minimum power transmission. The design goal is to jointly optimize the power allocation $\hat{p}_m$, receiving BF vectors $\hat{\mb{u}}^t_m$, and BS association to satisfy the SINR constraints. This leads to an uplink optimization problem described as
\begin{subequations} \label{obj_dual}
		\begin{eqnarray} 
 \hspace{-10mm} &\min \limits_{\lbrace \hat{\mb{u}}^t_m \rbrace, \lbrace \hat{p}_m \rbrace} & \hspace{-2mm} {} \scaleobj{.8}{\sum \limits_{m \in \mathcal{M}}} \hat{p}_m   \\
\hspace{-10mm} & \text{s.t.} & \hspace{-5mm}
 \max \limits_{t \in \mathcal{S}_m} \scaleobj{.8}{ \dfrac{\hat{p}_m |\hat{\mb{u}}^{t,H}_m \mb{g}^t_{m,m}|^2}{\sum \limits_{j \neq m} \hat{p}_j |\hat{\mb{u}}^{t,H}_m \mb{g}^t_{m,j}|^2 + \sigma^2 \hat{\mb{u}}^{t,H}_m \hat{\mb{u}}^t_m}} \geq \bar{\gamma}_m, \forall m \in \mathcal{M}.
\end{eqnarray} 
\end{subequations}

\begin{proposition} \label{up_down_prob}
Problem $(\mathcal{P}_3)$ can be solved via a dual uplink problem \eqref{obj_dual} with similar SINR constraints. Specifically, its Lagrangian dual
problem $(\mathcal{P}^{\sf{D}}_3)$ can be rewritten as
\begin{subequations} \label{obj_dual2}
		\begin{eqnarray} 
 \hspace{-12mm} &\min \limits_{\lbrace \hat{\mb{u}}^t_m \rbrace, \lbrace \lambda_m \rbrace} & \hspace{-2mm} {} \scaleobj{.8}{\sum \limits_{m \in \mathcal{M}}} \lambda_m \sigma^2   \\
\hspace{-12mm} & \text{s.t.} & \hspace{-7mm}
 \max \limits_{t \in \mathcal{S}_m} \scaleobj{.8}{ \dfrac{\lambda_m \sigma^2 |\hat{\mb{u}}^{t,H}_m \mb{g}^t_{m,m}|^2}{\sum \limits_{j \neq m} \lambda_j \sigma^2 |\hat{\mb{u}}^{t,H}_m \mb{g}^t_{m,j}|^2 + \sigma^2 \hat{\mb{u}}^{t,H}_m \hat{\mb{u}}^t_m}} \geq \bar{\gamma}_m, \forall m \in \mathcal{M}. \label{cnt_dual2}
\end{eqnarray}
\end{subequations}
where $\hat{p}_m = \lambda_m \sigma^2$. 
If \eqref{obj_dual2} is feasible, its optimal solution is also
optimal to $(\mathcal{P}^{\sf{D}}_3)$ . Otherwise, $(\mathcal{P}^{\sf{D}}_3)$  is unbounded above.
\end{proposition} 
\begin{IEEEproof}
    The proposition can be proved by following a similar
approach as in \cite{DuyNguyen_TWC17}. It can be briefly given as follows.
For given $\bs{\lambda}$, the optimal receive precoding vector at cluster $\mathcal{C}^t_m$ can be defined based on the MMSE receiver, i.e.,
\beq \label{LP_MMSE}
\hat{\mb{u}}^t_m = \Big( \scaleobj{.8}{\sum \limits_{\forall j}} \lambda_j \mb{g}^t_{m,j}\mb{g}^{t,H}_{m,j} + \mb{I}\Big)^{-1} \mb{g}^t_{m,m}.
\eeq
Substituting this result into constraint \eqref{cnt_dual2} yields, 
\beq \label{eq:dual3}
\lambda_m\Big(1+\scaleobj{.8}{\dfrac{1}{\bar{\gamma}_m}}\Big) \max \limits_{t \in \mathcal{S}_m}  \mb{g}^{t,H}_{m,m}\Big( \scaleobj{.8}{\sum \limits_{\forall j}} \lambda_j \mb{g}^t_{m,j}\mb{g}^{t,H}_{m,j} + \mb{I}\Big)^{-1} \mb{g}^t_{m,m} \geq 1.
\eeq
As can be observed, if \eqref{obj_dual2} is feasible, it is clear
that at optimality the set of inequality constraints \eqref{eq:dual3} must be
met at equality.
Regarding Lagrangian dual problem $(\mathcal{P}^{\sf{D}}_3)$, thanks to Lemma 1 in \cite{WeiYu_TSP07}, constraints \eqref{cnt_C5} can be recast as,
\beq \label{eq:dual4}
\lambda_m\Big(1+\scaleobj{.8}{\dfrac{1}{\bar{\gamma}_m}}\Big) \max \limits_{t \in \mathcal{S}_m}  \mb{g}^{t,H}_{m,m}\Big( \scaleobj{.8}{\sum \limits_{\forall j}} \lambda_j \mb{g}^t_{m,j}\mb{g}^{t,H}_{m,j} + \mb{I}\Big)^{-1} \mb{g}^t_{m,m} \leq 1.
\eeq
Again, if $(\mathcal{P}^{\sf{D}}_3)$ is feasible, these constraints \eqref{eq:dual4} should be also met at equality.
Therefore, problems \eqref{eq:dual3} and $(\mathcal{P}^{\sf{D}}_3)$ are equivalent since $\lambda_m$'s in both problems are the fixed point of the following equations,
\beq \label{eq:dual5}
\lambda_m\Big(1+\scaleobj{.8}{\dfrac{1}{\bar{\gamma}_m}}\Big) \max \limits_{t \in \mathcal{S}_m}  \mb{g}^{t,H}_{m,m}\Big( \scaleobj{.8}{\sum \limits_{\forall j}} \lambda_j \mb{g}^t_{m,j}\mb{g}^{t,H}_{m,j} + \mb{I}\Big)^{-1} \mb{g}^t_{m,m} = 1.
\eeq
The results given in \cite{DuyNguyen_TWC17,WeiYu_TSP07} have proved that this fixed point is unique if it
exists. Under such circumstances, this fixed point represents the optimal solution to both problems.
\end{IEEEproof}
\subsection{Iterative Algorithm for Solving Problem $(\mathcal{P}_3)$}
This section focuses on identifying the fixed point for equations \eqref{eq:dual5}, from which the optimal LP vectors can be established. By reformulating \eqref{eq:dual5} into a fixed point iteration format, $\lambda_m$ in the $(q+1)^{th}$ iteration can be obtained as
\beq \label{iter_process}
\lambda_m^{(q+1)} = \min \limits_{t \in \mathcal{S}_m} f^t_m(\bs{\lambda}^{(q)}),
\eeq
where $f^t_m(\bs{\lambda}) = (1+1/\bar{\gamma}_m)^{-1}/\big[ \mb{g}^{t,H}_{m,m}\big( \sum_{\forall j} \lambda_j \mb{g}^t_{j,m}\mb{g}^{t,H}_{j,m} + \mb{I}\big)^{-1} \mb{g}^t_{m,m}\big]$. Due to \cite{DuyNguyen_TWC17,WeiYu_TSP07}, $f^t_m(\bs{\lambda})$ qualifies as a standard function, adhering to three key properties: positivity, monotonicity, and scalability. Consequently, if a fixed point of \eqref{eq:dual5} exists, its uniqueness is guaranteed, and the iterative process detailed in \eqref{iter_process} will converge fast to the fixed point.

Once, $\bs{\lambda}^{\star}$ is determined, according to that given in the proof of Theorem~\ref{thm_duality}, the cluster serving user $m$ can be defined as 
\beq
t_m = \argmin_{t \in \mathcal{S}_m} f^t_m(\bs{\lambda}^{\star}).
\eeq
Then, we set $\mathbf{u}_{m}^{t \star}$ to all-0 vector for all $t \in \mathcal{S}_m/\{t_m\}$. Utilizing $\hat{\mb{u}}^{t_m}_m $ given in \eqref{LP_MMSE} to indicate $\mathbf{u}_{m}^{t_m \star}$, we substitute $\mathbf{u}_{m}^{t_m \star} = \sqrt{\delta_m} \hat{\mb{u}}^{t_m}_m$ into \eqref{lagrangian_cnt} to obtain the following equation,
\beq \label{eq:delta1}
\dfrac{\delta_m}{\bar{\gamma}_m}|\mb{g}^{t_m,H}_{m,m} \hat{\mb{u}}^{t_m}_m|^2 - \delta_j \scaleobj{.8}{\sum \limits_{j \neq m}} |\mb{g}^{t_j,H}_{j,m} \hat{\mb{u}}^{t_j}_j|^2 = \sigma^2, \; \forall m.
\eeq
Defining $\mb{F} \in \mathbb{R}^{M \times M}$ with $[\mb{F}]_{m,m}=(1/\bar{\gamma}_m)|\mb{g}^{t_m,H}_{m,m} \hat{\mb{u}}^{t_m}_m|^2$ and 
$[\mb{F}]_{m,j}=-|\mb{g}^{t_j,H}_{m,j} \hat{\mb{u}}^{t_j}_j|^2$, \eqref{eq:delta1} can be rewritten as $\mb{F}\bs{\delta} = \mb{1}\sigma^2$ where $\bs{\delta} = [\delta_1,...,\delta_M]^T$. Based on that, $\bs{\delta}$ can be defined as
\beq \label{eq:delta}
\bs{\delta} = \mb{F}^{-1} \mb{1}\sigma^2.
\eeq
These processes are summarized in Algorithm~\ref{alg:gms1}. 
It is worth noting that Algorithm~\ref{alg:gms1} returns the optimal solution of problem~\eqref{obj_dual2} which is also optimal to $(\mathcal{P}_3^{D})$ and $(\mathcal{P}_3)$, thanks to Theorem~\ref{thm_duality} and Proposition~\ref{up_down_prob}.
Furthermore, this optimal solution satisfies \textbf{Cond-1}; hence, this is the solution of $(\mathcal{P}_2)$. 

\begin{algorithm}[!t]
\caption{\footnotesize\textsc{Duality Cluster-Association and LP Design}}
\label{alg:gms1}
\footnotesize
\begin{algorithmic}[1]
\STATE Initialize $\lambda_m > 0, \; \forall m \in \mathcal{M}$.
\REPEAT
\STATE Update $\lambda_m$ as described in \eqref{iter_process}.
\UNTIL{Convergence to $\bs{\lambda}^{\star}$.}
\FOR{$m=1:M$} 
\STATE Select cluster $t_m = \argmin_{t \in \mathcal{S}_m} f^t_m(\bs{\lambda}^{\star})$.
\STATE Set $\mathbf{u}_{m}^{t \star}$ to all-0 vector for all $t \in \mathcal{S}_m/\{t_m\}$.
\ENDFOR
\STATE Calculate $\bs{\delta}$ as described in \eqref{eq:delta}.
\STATE Determine the LP vector for user $m$ as in $\mathbf{u}_{m}^{t_m \star} = \sqrt{\delta_m} \hat{\mb{u}}^{t_m}_m$, for all $m \in \mathcal{M}$.
\end{algorithmic}
\normalsize
\end{algorithm}

\subsection{Simple Solution and Complexity Analysis}
\subsubsection{Simple Solution}For comparison purposes, this section introduces a simple solution where the CA process can be simplified based on the effective channel vectors.
Specifically, the cluster solution for user $m$ can be defined as $t_m^{\prime} = \argmax_{t \in \mathcal{S}_m} |\mb{g}^t_{m,m}|^2$.
Then, the corresponding LP vectors can be determined by exploiting the uplink-downlink duality approach given in \cite{WeiYu_TSP07} which is similar to the design process presented in the previous section. In particular, the simple solution is summarized in Algorithm~\ref{alg:gms2}.


\begin{algorithm}[!t]
\caption{\footnotesize\textsc{Simple Algorithm}}
\label{alg:gms2}
\footnotesize
\begin{algorithmic}[1]
\STATE For every user $m$, define the associated cluster as $t_m^{\prime} = \argmax_{t \in \mathcal{S}_m} |\mb{g}^t_{m,m}|^2$.
\REPEAT
\STATE Update $\lambda_m$ as $\lambda_m^{(q+1)} =  f^{t^{\prime}_m}_m(\bs{\lambda}^{(q)}),$.
\UNTIL{Convergence to $\bs{\lambda}^{\prime}$.}
\STATE Calculate $\bs{\delta}^{\prime}$ as described in \eqref{eq:delta} with $\bs{\lambda}^{\prime}$ and $\{t^{\prime}_m\}$'s.
\STATE For every user $m$, determine its LP vector $\mathbf{u}_{m}^{t^{\prime}_m} = \sqrt{\delta^{\prime}_m} \hat{\mb{u}}^{t^{\prime}_m}_m$, for all $m \in \mathcal{M}$. Set $\mathbf{u}_{m}^{t}$ to all-0 vector for all $t \in \mathcal{S}_m/\{t^{\prime}_m\}$.
\end{algorithmic}
\normalsize
\end{algorithm}
\subsubsection{Complexity Analysis}

This section investigates the complexities of our two proposed approaches. We observe that both algorithms comprise two processes: (i) iterative updating of $\boldsymbol{\lambda}$ and (ii) determining LP vectors. 
While the second process remains consistent across both approaches, a notable difference exists in their implementation for the first. Specifically, during each iteration of updating $\boldsymbol{\lambda}$, for each user ($m$), Algorithm~\ref{alg:gms1} calculates the value of $f^t_m(\boldsymbol{\lambda})$ function $T_m$ times to determine $t_m$, whereas Algorithm~\ref{alg:gms2} only requires a single calculation of $f^{t'}m(\boldsymbol{\lambda})$. Let $J_{\sf{it}}$ represent the average number of iterations for the first process in both algorithms. Research in \cite{DuyNguyen_TWC17,WeiYu_TSP07} confirms that the iterative process converges rapidly when a fixed point exists. Considering the computational efforts related to matrix inversion and multiplication, the complexities of our two proposed approaches are presented as follows:
\vspace{-2mm}
\beqn
X_{\sf{Alg.1}} &=& \mathcal{O}(J_{\sf{it}} L M S^B B^3 + M^3), \\
X_{\sf{Alg.2}} &=& \mathcal{O}(LMB^2 + J_{\sf{it}}MB^3+ M^3).
\eeqn
where $S$ represents the maximum element number of $\mathcal{B}^{\ell}_m$'s.

\section{Numerical Results}
\vspace{-1mm}
 \begin{figure}[!t]
    \centering
    \includegraphics[width=50mm]{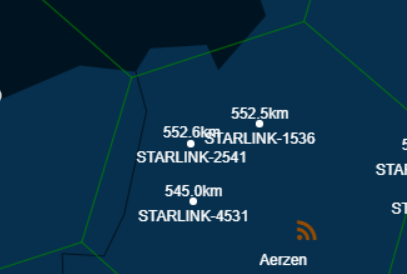}
    \vspace{-3mm}
    \caption{LEO-position simulation settings -``https://satellitemap.space/''.}
    \label{fig:simluation_Sat}
    \vspace{-3mm}
\end{figure}

\begin{table}[!t]
\footnotesize
	\centering
	\caption{\textsc{Simulation Parameters}}
	\label{tab:simpara}
 \vspace{-2mm}
	\begin{tabular}{l | r}
		\toprule
		\midrule
		Forward link carrier frequency										& $19$~GHz \\
		Number of simulated users					& $10-70$ \\
		Uniform rectangular array (URA) size							& $10\times10$\\
		Array element normalized spacing &  $2.5$ \\
		Array element radiation model & $2\times2$ URA sub-array \\ 
		FFT size & $16 \times 16$ (2D)\\
		User terminal antenna gain												& $41.45$~dBi\\
		Temperature at user terminals													& $224.5$~K\\
		Channel Model (Adopt and calibrate for LEO)						&  Refer to~\cite{Angeletti_Access20}   \\
		\bottomrule
	\end{tabular}
 \vspace{-3mm}
	\end{table}

\begin{figure}[!t]
    \centering
    \includegraphics[width=85mm]{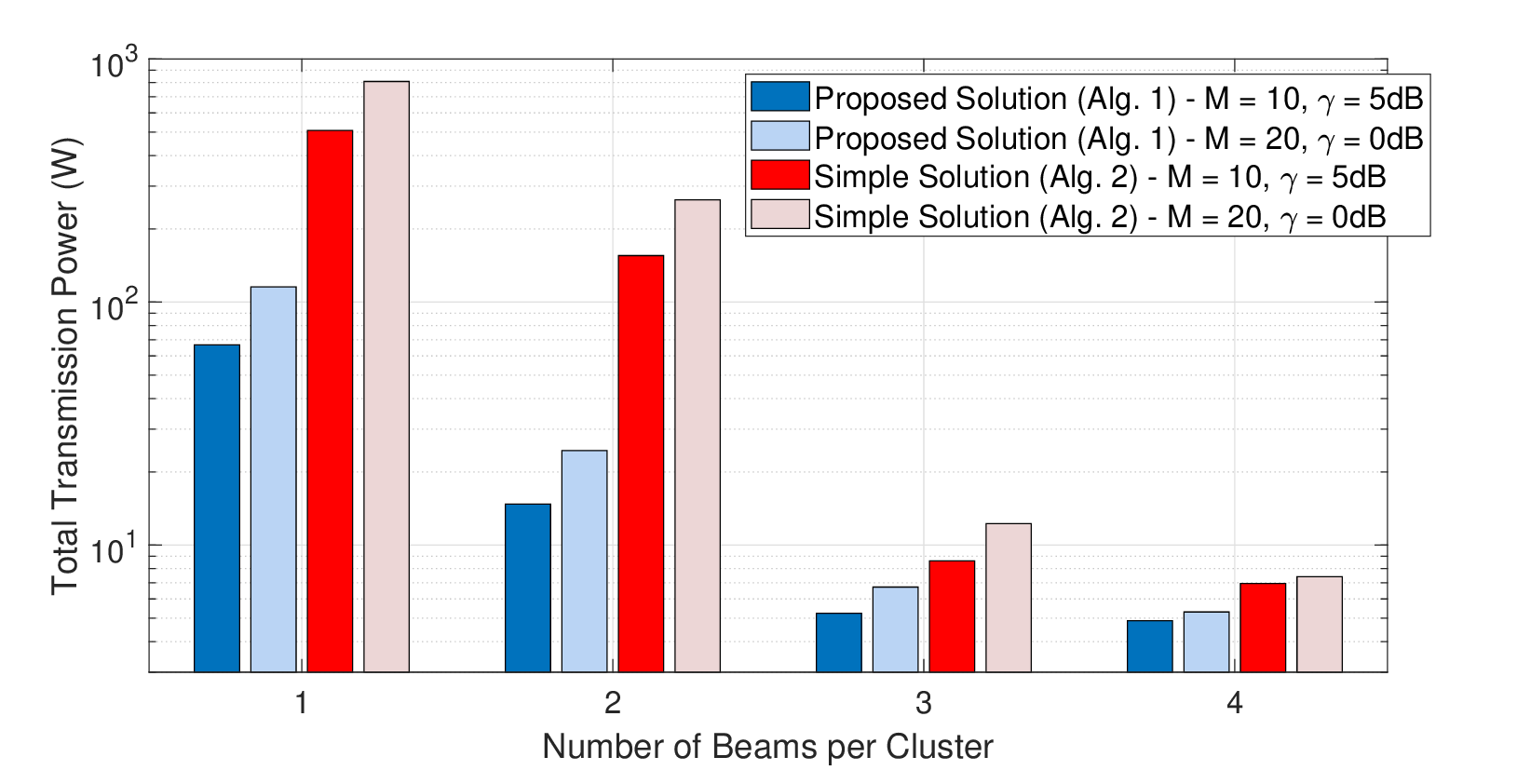}
    \vspace{-3mm}
    \caption{Total transmission power vs cluster sizes.}
    \label{fig:P_Vs_B}
    \vspace{-3mm}
\end{figure}

In this simulation, we analyze an MSat system composed of three STARLINK LEOSats with locations (Latitude, Longitude) at $(52.817247, 9.291984)$, $(52.589261, 7.669242)$, and $(52.054784, 7.876349)$. We consider several users randomly distributed within a rectangular area defined by latitude and longitude limits from $(51.0, 5.5)$ to $(54.0, 9.5)$. The simulation parameters are detailed in Table~\ref{tab:simpara}. Channel vectors from the LEOSats to users are generated based on the model described in \cite{Angeletti_Access20} with movement-regarded calibration, while the DFT BF vectors are adopted from our previous research \cite{VuHa_GCWS22,Rakesh_PIMRC23}. To assign $\mathcal{B}^{\ell}_m$ for user $m$ (for any $\ell$ and $m$), we calculate its $(U,V)$ coordinate from the perspective of LEOSat $\ell$, compare this coordinate with those of the beam centers illustrated in Fig.~\ref{fig:SNR8x8taper}, and identify the five nearest beams as the elements of $\mathcal{B}^{\ell}_m$. Then, we fix the cluster size at $3$, i.e., $B = 3$ except in the simulation relating to Fig.~\ref{fig:P_Vs_B}.
The target SINRs of all users are set the same as $\gamma$, i.e., $\bar{\gamma}_m = \gamma, \; \forall m \in \mathcal{M}$, and the value of $\gamma$ is varied in different simulation settings.

In Fig.~\ref{fig:P_Vs_B}, we present the total transmission power returned by our proposed methods as the cluster size ($B$) varies, to assess the impact of MB LP design on SATCOM performance. 
Here, a cluster size of $1$ implies that only a single beam from the associated LEOSat is selected to serve each user.
As expected, both algorithms exhibit a reduction in transmission power with increasing cluster sizes. 
Specifically, with the setting of $M=10$ and $\gamma = 5$~dB, the transmission power according to Algorithm~\ref{alg:gms1} decreases approximately five-fold when the cluster size increases from $1$ to $2$. The total required power can be degraded by $3$ times further when the cluster size expands to $3$, before seemingly saturating at a cluster size of $4$. 
Similar trends can be observed for the outputs of Algorithm~\ref{alg:gms1} under a different simulation setting and also for Algorithm~\ref{alg:gms2} across all settings. 
These results significantly highlight the advantages of LP technologies in MB-enabled SATCOM systems. Notably, Algorithm~\ref{alg:gms1} consistently requires less transmission power than Algorithm~\ref{alg:gms2}, showcasing the effectiveness of our joint LP and CA strategy in managing MB transmission of the CoMSat systems.

 \begin{figure}[!t]
    \centering
    \includegraphics[width=85mm]{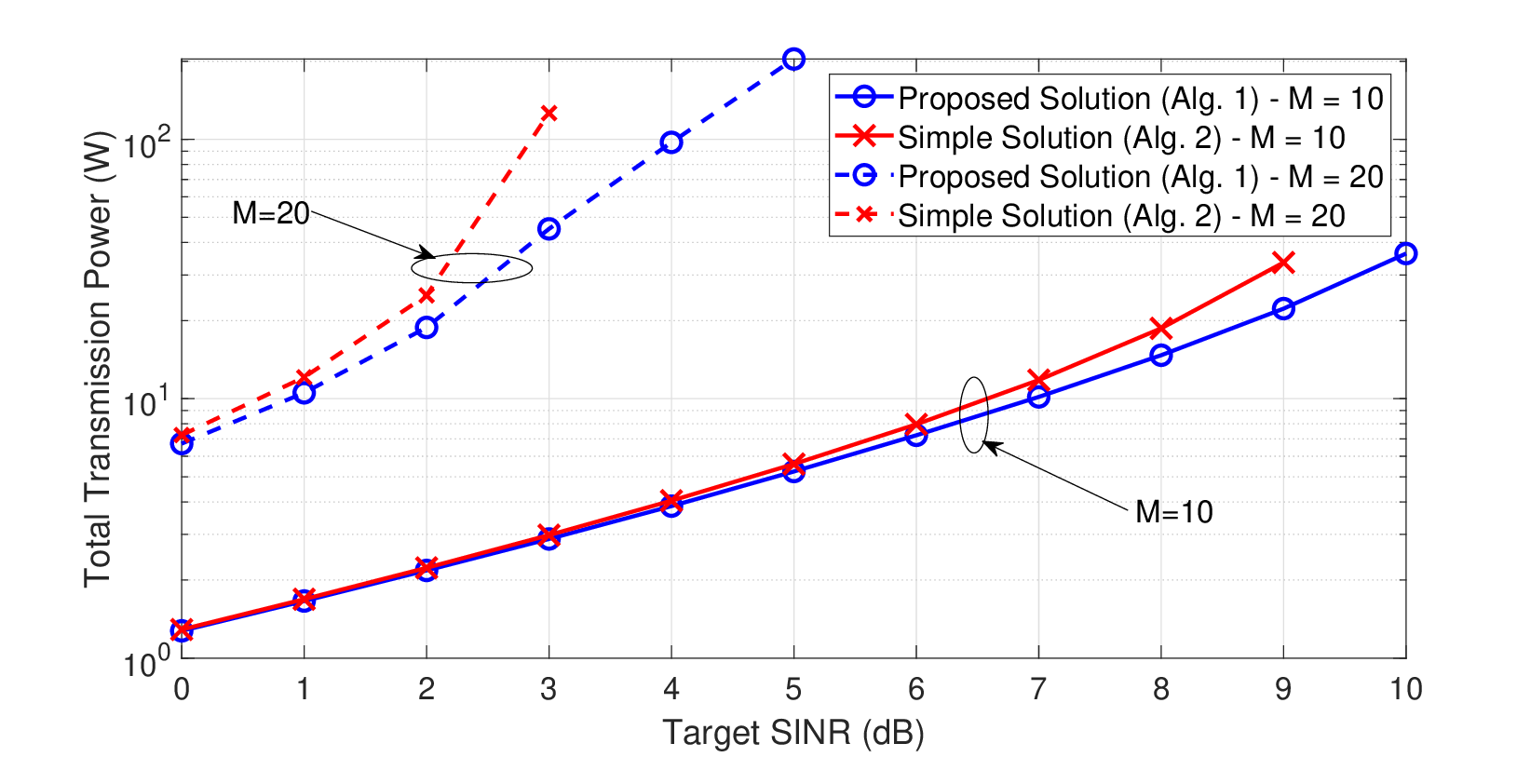}
    \vspace{-3mm}
    \caption{Total transmission power vs target SINRs.}
    \label{fig:P_Vs_SINR}
    \vspace{-3mm}
\end{figure}

 \begin{figure}[!t]
    \centering
    \includegraphics[width=85mm]{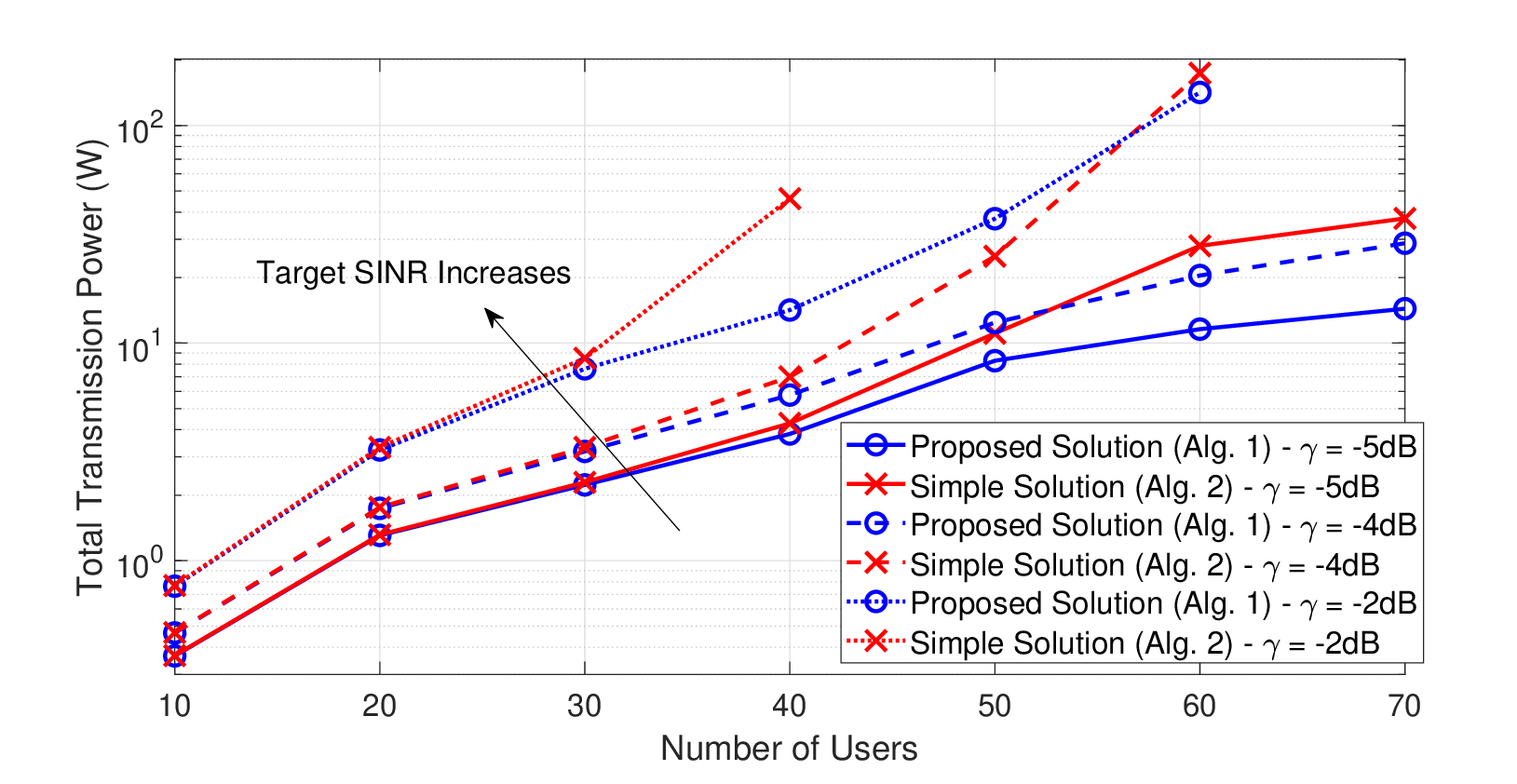}
    \vspace{-3mm}
    \caption{Total transmission power vs numbers of users.}
    \label{fig:P_Vs_M}
    \vspace{-3mm}
\end{figure}

Figs.~\ref{fig:P_Vs_SINR} and \ref{fig:P_Vs_M} show how the total transmission power required by the two proposed approaches varies with different target SINR values for users and the number of users in the system under various simulation settings. As anticipated, the need for higher SINRs for users leads to an increased demand for transmission energy to meet these targets. Furthermore, the system requires more transmission power as the number of users increases. Again, Algorithm~\ref{alg:gms1} consistently returns lower required transmission power than Algorithm~\ref{alg:gms2} does. Notably, as the system's transmission load increases, either by raising the target SINR or by adding more users to the network, the difference in power requirement between the two algorithms grows exponentially.
These findings distinctly highlight the advantages of using a jointly designed LP and CA mechanism.

\section{Conclusion} \label{ccls}
\vspace{-1mm}

This paper has presented a novel joint optimization framework enhancing MB transmission and CoMSat systems via efficient CA, beam selection, and LP designs. Utilizing downlink-uplink duality, we devised an efficient iterative method for optimal solutions and introduced a simpler alternative for beam selection and LP design. Our simulations confirm the effectiveness and adaptability of these approaches across different settings, contributing theoretical insights and practical strategies to improve SATCOM performance.

\section*{Acknowledgment}
\footnotesize
This work was supported in part by European Space Agency (ESA) under EGERTON project (4000134678/21/UK/AL). The opinions, interpretations, recommendations, and conclusions presented in this paper are those of the authors and are not necessarily endorsed by ESA. This work was also funded in part by the Luxembourg National Research Fund (FNR), with two granted projects corresponding to two grant references
C23/IS/18073708/SENTRY (SENTRY project) and C21/IS/16352790/ARMMONY (ARMMONY project).
	
\bibliographystyle{IEEEtran}
\bibliography{reference}

\end{document}